\documentclass[journal=jacsat,manuscript=article]{achemso}

\usepackage[version=3]{mhchem} 

\author{J\'er\^ome Tribollet}
\email{tribollet@unistra.fr}
\affiliation[Institut de Chimie]
{Institut de Chimie de Strasbourg, Strasbourg University, UMR 7177 (CNRS-UDS),\\
4 rue Blaise Pascal, CS 90032, F-67081 Strasbourg Cedex, 
France} 

\title[An \textsf{achemso} demo]
  {Hybrid nanophotonic-nanomagnonic\\SiC-YiG quantum sensor:\\ II/ optical fiber based ODMR and OP-PELDOR\\ experiments on bulk HPSI 4H-SiC.}

\abbreviations{EPR (Electron Paramagnetic Resonance)} 
  

\begin{document}
\begin{abstract}
Here I present my first fiber based coupled optical and EPR experiments associated to the development of a new SiC-YiG quantum sensor that I recently theoretically described (arxiv:1912.11634). This quantum sensor was designed to allow sub-nanoscale single external spin sensitivity optically detected pulsed electron electron double resonance spectroscopy, using an X band pulsed EPR spectrometer, an optical fiber, and a photoluminescence setup. First key experiments before the demonstration of ODPELDOR spectroscopy are presented here. They were performed on a bulk 4H-SiC sample containing an ensemble of residual V2 color centers (spin S=3/2). Here I demonstrate i/ optical pumping assisted pulsed EPR experiments, ii/ fiber based ODMR and optically detected RABI oscillations, and iii/ optical pumping assisted PELDOR experiments, and iv/ some spin wave resonance experiments. Those experiments confirm the feasability of the new quantum sensing approach proposed.
\end{abstract} 

\section{Introduction}  
I recently presented the theory~\cite{Tribollet2019dec} of a new SiC-YiG quantum sensor and the associated state of art optically detected pulsed double electron electron spin resonance spectroscopy (OD-PELDOR), allowing sub-nanoscale single external spin sensing. This new methodology requires only the use of a standard X band pulsed EPR spectrometer~\cite{Schweiger2001}, as well as  an optical fiber and a new SiC-YiG quantum sensor. The fiber and the quantum sensor can be both introduced in a standard EPR tube.\\ Here I present my first combined pulsed EPR and optical experiments, all performed on a commercially available bulk 4H-SiC HPSI sample, naturally containing a diluted ensemble of V2 color centers spin  probes~\cite{Kraus2014,Widmann2015,Baranov2013,Fuchs2017,Tarasenko2018}. The aim of those first experiments is to demonstrate the relevance and feasability of interfacing a standard optical setup for photoluminescence excitation and collection with a commercial pulsed EPR/ pulsed ELDOR ELEXYS E 580 spectrometer operating at X band from Bruker, by means of a single optical fiber (or a fiber bundle). This setup allows to perform ODMR and optical pumping assisted PELDOR (puulsed electron electron double resonance) experiments~\cite{Schweiger2001}, which are key intermediate experiments to perform, before the demonstration of pulsed ODPELDOR experiments with a SiC-YiG quantum sensor. The whole experimental setup I used corresponds to the one described on fig.2 of my previous theoretical work~\cite{Tribollet2019dec}, the coupler between the SiC sample and the optical fiber being here a GRIN microlens. The optical fiber, the GRIN microlens and the 4H-SiC sample are all introduced in an EPR tube, which itself is inserted inside the pulsed EPR resonator, a flexline resonator from Bruker (MD5 or MS3 depending on experiments).  The pulsed EPR resonator itself is introduced inside an Oxford CF935 continuous flow cryostat for pulsed EPR spectroscopy at variable temperature (4K-300K). When necessary, a 785 nm laser was used for optical pumping of V2 spins and for optical excitation of the V2 color center photoluminescence, centered around 915 nm at low temperature. This photoluminescence was detected, after optical filtering, by a silicon photodiode, in all presented ODMR experiments. Excitation and collection of the photoluminescence of the SiC sample was performed using the same optical fiber by means of a dichroic mirror. A lock in amplifier or a transient recorder were used for data acquisition, which were visualized on the XEPR software of Bruker provided with the ELEXYS E580 pulsed EPR spectrometer.\\   

\section{Optical pumping assisted EPR and ODMR characterization of V2 spins in bulk 4H-SiC}    
 
First, I demonstrate on fig.1 that the Electron Paramagnetic Resonance (EPR) rotational pattern of the V2 color centers spins in bulk 4H-SiC can be recorded, under optical pumping conditions with this experimental setup, allowing to check the zero field splitting and g factor of those paramagnetic centers~\cite{Kraus2014,Widmann2015,Baranov2013,Fuchs2017,Tarasenko2018}, and finally to identify them.    
 
\begin{figure}[ht]  
\centering \includegraphics[width=1.1 \textwidth]{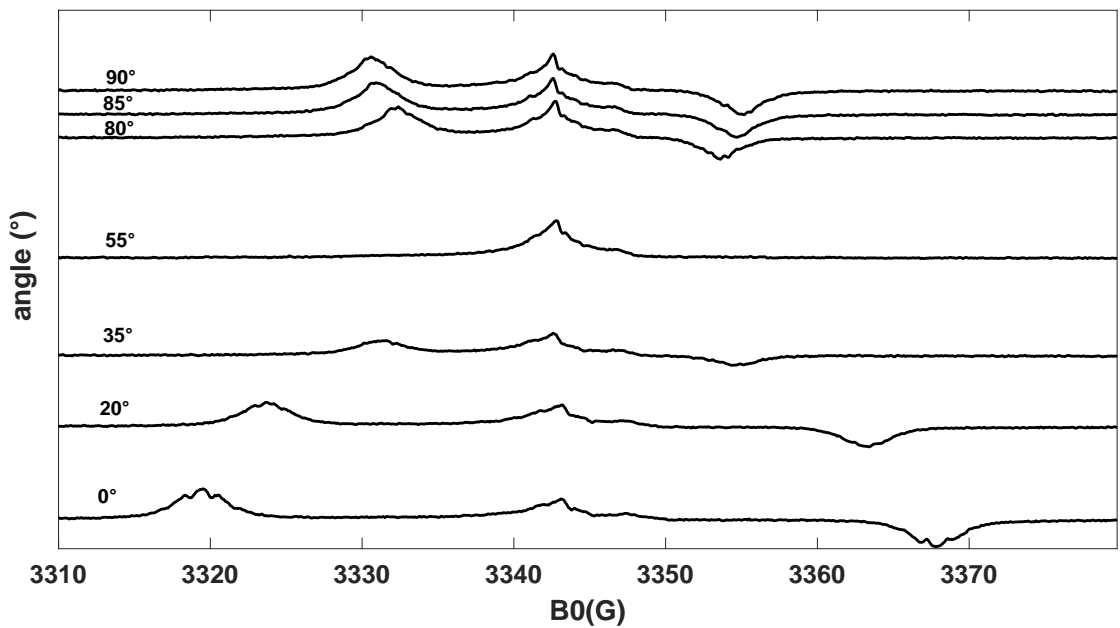}
\caption{\label{fig_10} CW EPR rotational pattern of the V2 spins in bulk HPSI 4H-SiC recorded at room temperature and at X band (f=9.369 GHz) under continuous optical pumping, with a laser at 785 nm providing a power of P=39 mW at the outpout of optical fiber. The nul angle correspond to the external magnetic field aligned along the c axis of 4H-SiC.} 
\end{figure} 

This rotational pattern was obtained here at X band and room temperature, using cw EPR under continuous optical pumping with a 785 nm laser. The optical pumping effect is clearly seen on the shape of the EPR spectrum of fig.1: the left side positive signal correspond to an EPR transition with induced absorption, while the negative one on right side corresponds to stimulated emission associated to population inversion on this EPR transition. The rotational pattern of fig.1 can be well reproduced (except the optical pumping effects) by a numerical simulation with Easyspin~\cite{Stoll2006}, as shown on fig.2, considering a spin S=3/2 with a zero field splitting D = 35 MHz and an isotropic g factor g = 2.0028, confirming previously obtained magnetic parameters of the V2 spin hamiltonian in 4H-SiC~\cite{Kraus2014,Widmann2015,Baranov2013,Fuchs2017,Tarasenko2018}. 

\begin{figure*}[ht]  
\centering \includegraphics[width=1.2 \textwidth]{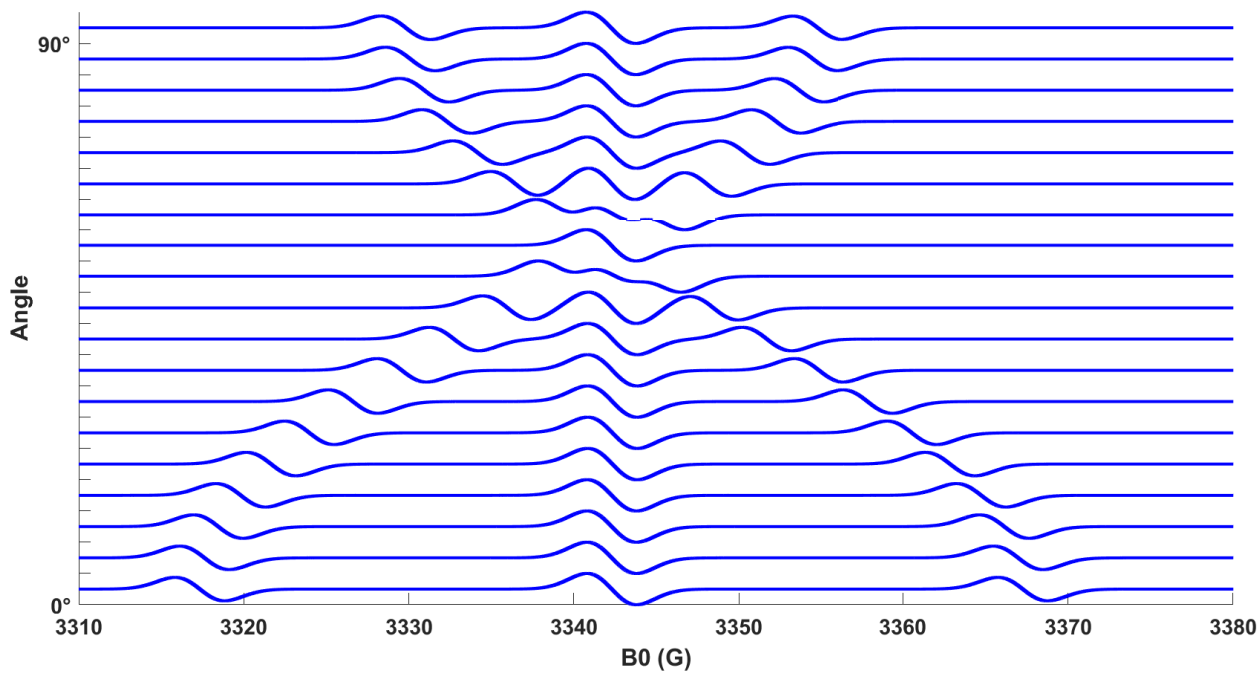}
\caption{\label{fig_12} Numerical simulation with Easyspin of the V2 rotational pattern in 4H-SiC, considering a spin S=3/2 with a zero field splitting D= 35 MHz and an isotropic g factor g=2.0028; f=9.369 GHz; the linewidth is 3G here. The nul angle correspond to the external magnetic field aligned along the c axis of 4H-SiC. Derivative of absorption curves are shown here. The spin state populations assumed here are those of the thermal equilibrium.} 
\end{figure*} 

Secondly, I demonstrate on fig.3, that several EPR experiments on V2 color centers spins in bulk 4H-SiC and under continuous optical pumping are possible with this fiber based ODMR setup, that is, from top to bottom, room temperature cw EPR, room temperature pulsed EPR, room temperature ODMR, and 90K ODMR. The two first experiments benefits from the fact that a large ensemble of V2 spins is present in this 4H SiC HPSI sample allowing a direct detection of the EPR signal here, without any use of the photoluminescence, the optical fiber being however used for optical pumping of the V2 spins. However, the last two experiments presented on fig.3 are true ODMR experiments, meaning that the photoluminescence of the V2 spin probe is the recorded signal along the vertical axis of those two ODMR experiments. 

\begin{figure}[ht] 
\centering \includegraphics[width=1.1 \textwidth]{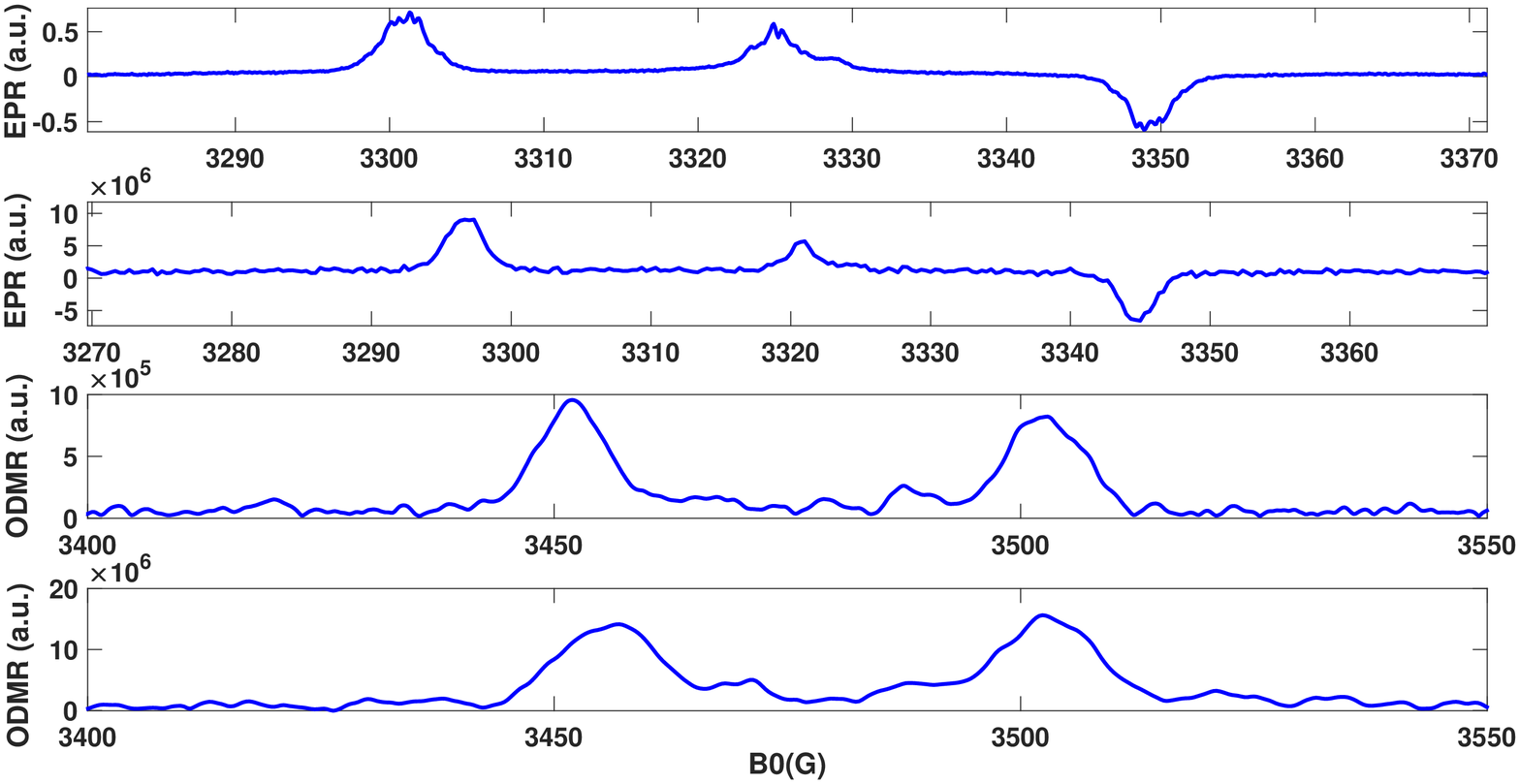}
\caption{\label{fig_10} From top to bottom and under continuous optical pumping at 785 nm: a/room temperature cw EPR spectrum (f=9.320 GHz, MS3, 36mW at 785 nm), b/ room temperature field sweep pulsed EPR spectrum (f=9.308 GHz, MS3, 36 mW at 785 nm, recorded at $2\:\tau\:\approx\:2.4\:\mu\!s$), c/ room temperature ODMR spectrum (f=9.743 GHz, MD5, 36 mW at 785 nm), and d/ 90K ODMR spectrum (f=9.746 GHz, MD5, 30 mW at 785 nm). The static magnetic field B0 is applied along the c axis of 4H-SiC.} 
\end{figure} 

Note that the two ODMR spectrum of the V2 spins in 4H SiC presented here have been obtained under continuous optical pumping. To sensitively detect the ODMR spectrum, a train of periodic microwave pulses is send on the sample, such that instead of having a constant rate of photoluminescence under continuous optical excitation, one obtains a periodic modulation of the photoluminescence signal following the period of the microwave pulses, but only when a paramagnetic resonance of the V2 spins is excited by the microwaves. 

\begin{figure}[ht] 
\centering \includegraphics[width=1.1 \textwidth]{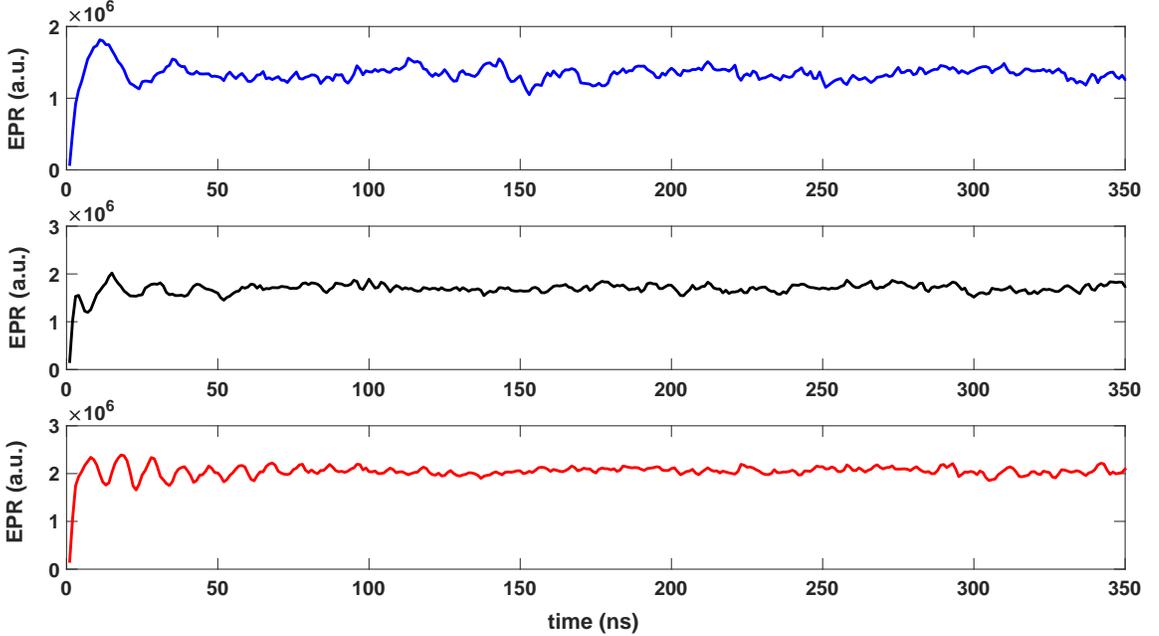}
\caption{\label{fig_10} ODMR detected coherent RABI oscillations of V2 spins in 4H SiC recorded at different microwave power attenuation (from top to bottom: 10 dB, 5 dB, 0 dB attenuation, MD5, B0= 3457 G, f= 9.746 GHz), for various duration of the nutation microwave pulse applied (time). Periodically cycling this experiment, again allows to obtain a modulated amount of photoluminescence easily detected by a lock in amplifier. Those curves demonstrate the quantum coherent control of V2 spins by microwave pulses (nutation or single qubit quantum gate performed over a spin ensemble) and also their optical detection by photoluminescence, in a single ODMR nutation experiment.} 
\end{figure} 

Such paramagnetic resonance induces a change of spin state populations in the ground state, further converted in a change of the amount of photoluminescence collected under optical excitation. Such periodic photoluminescence signal is then collected by a photodiode and then send into a lock in amplifier allowing an efficient extraction of the amount of photoluminescence modulated at a frequency inversely proportionnal to the period between two successive microwave Pi pulses applied on V2 spins. Note that EPR nutation experiments, also called RABI oscillations measurements, were generally performed on V2 spins ensemble under optical pumping before such experiments, either using direct detection (not shown) or photoluminescence detection of the RABI oscillation as shown on fig.4, in order to check the appropriate microwave Pi pulse parameters (choosing microwave power and pulse duration). Once the train of microwave Pi pulses resonant with the V2 spins has been adjusted, the ODMR measurement is easy to perform with the lock in amplifier.

\section{Spin decoherence and spin relaxation time of bulk V2 spins}
Thirdly, I demonstrate on fig.5 how to measure at room temperature the spin coherence time T2, the longitudinal relaxation time T1 and the optical pumping time Top of the bulk V2 spins with appropriate experiments combining an optical pumping pulse and appropriately synchronized and time delayed microwave pulses. At room temperature and with those optical pumping conditions, I find $T_{op}\:=\:139\:\mu\!s$, $T_{1}\:=\:354\:\mu\!s$, and $T_{2}\:=\:48\:\mu\!s$.
 
\begin{figure}[htb] 
\centering \includegraphics[width=0.8 \textwidth]{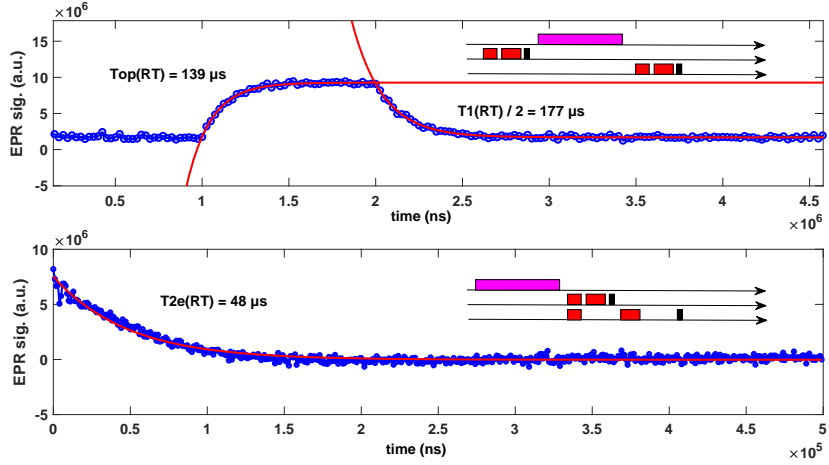}
\caption{\label{fig_10} Top curve: measurement of $T_{op}$ and $T_{1}$ of bulk V2 spins in 4H-SiC at room temperature. A long optical pumping pulse of 1 ms at 785 nm and 36 mw optical power at the fiber output is used and has a fixed temporal position in the sequence. A standard direct detection spin echo sequence ($\frac{\pi}{2}\:\tau\:\pi\:\tau\:echo$) is synchronized with the optical pumping pulse and globally translated in time through the optical pumping pulse, allowing to follow the time evolution of the spin state populations associated to a given EPR transition of the V2 spins (B0= 3297 G , f=9.308 GHz, MS3), before , during and after the optical pumping pulse. The data are in blue, the two monoexponential fit are in red, providing: $T_{op}\:=\:139\:\mu\!s$ and $T_{1}\:=\:354\:\mu\!s$. Bottom curve: Standard Hahn spin echo decay curve ($\frac{\pi}{2}\:\tau\:\pi\:\tau\:echo$) recorded on one transition of the V2 (B0= 3297 G , f=9.308 GHz, MS3), where the spin echo is recorded at various delays $2\:\tau$, the first microwave pulse starting $20\:\mu\!s$ after a long optical pumping pulse of $900\:\mu\!s$ at 785 nm and with 36 mw of optical power at the fiber output. The data are in blue, the monoexponential fit is in red, providing: $T_{2}\:=\:48\:\mu\!s$. Inset: pulses sequences applied: pink: optical pulse; red: microwave pulse; dark:spin echo for direct detection. The third arrow compared to second one shows which delay parameter vary.} 
\end{figure} 

Those values of $T_{1}$ and $T_{2}$ are in good agreement with values reported for bulk V2 spins in 4HSiC at room temperature~\cite{Kraus2014,Widmann2015,Baranov2013,Fuchs2017,Tarasenko2018}. One note that it was also previously reported that the optical pumping time $T_{op}$ depends on the optical power send on the V2 spins~\cite{Fischer2018} (and thus on the laser and coupler used) and on the temperature which controls $T_{1}$. One can also note here that for using optimally the SiC-YIG quantum sensor which I described in my previous theoretical work~\cite{Tribollet2019dec}, a spin coherence time for an isolated sub-surface V2 spin probe of $T_{2}\:=\:12.5\:\mu\!s$ was assumed. Thus, ODPELDOR spectroscopy should be feasible if the SiC surface defects can be made sufficiently silent, either by surface passivation or by cryogenic cooling of the whole quantum sensor, in order that the T2 of sub surface V2 has a value on the order of the one of bulk V2 spins found here.

\section{Optical pumping assisted PELDOR spectroscopy and quantum sensing of carbon related defects by bulk V2 spins}

Finally, the last key experiment combining optical and EPR tools which is demonstrated here and which allows to go a step further towards fiber based ODPELDOR quantum sensing with a SiC-YIG quantum sensor, is optical pumping assisted PELDOR experiment, as shown on fig.6. It is a pump-probe like two microwave frequencies experiment combined with optical pumping, as explained in my previous theoretical work~\cite{Tribollet2019dec}. It is conveniently implemented here by interfacing, through an optical fiber, the capabilities of the commercial pulsed ELDOR spectrometer (ELEXYS E580) and the ones of the outside optical setup. The two PELDOR experiments presented here are four-pulse DEER (double electron electron resonance) experiments~\cite{Schweiger2001} combined, either with a continuous optical pumping and direct EPR detection of the stimulated echo (top spectrum), or with a transient optical pumping pulse and a direct detection of the refocused echo (bottom spectrum). When the pump frequency fp is resonant with any spin specy present in the sample and physically close to the V2 spins probe, then a driven decoherence effect occurs producing a reduced spin echo of the V2 spins probes, and thus a dip in the PELDOR spetrum. As the stimulated echo has a larger amplitude than the refocused echo, the PELDOR spectrum of fig.6 a/ (top) has a better signal to noise ratio than the one of fig.6 b/ using the refocused echo, which is generally used in structural biology~\cite{Schweiger2001}. The comparison also shows that continuous optical pumping seems not to induce a decrease of the signal to noise ratio of such PELDOR experiment.

\begin{figure}[ht] 
\centering \includegraphics[width=1.1 \textwidth]{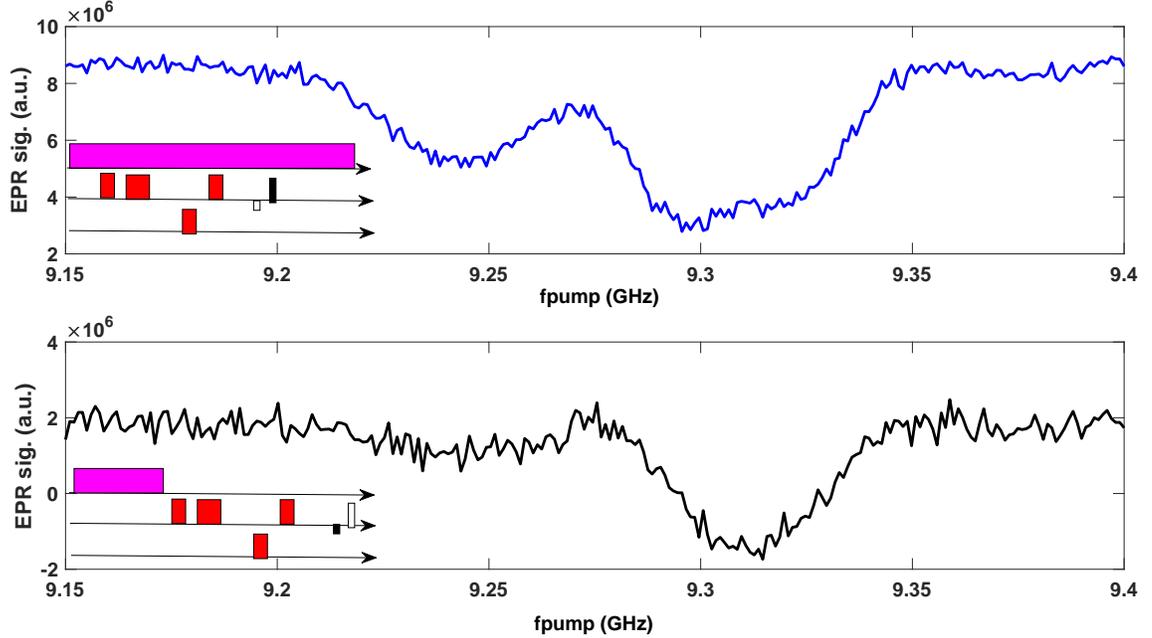}
\caption{\label{fig_10} PELDOR experiments of the type four-pulse DEER experiments combined, either with a continuous optical pumping and direct EPR detection of the stimulated echo (a/ top spectrum), or with a transient optical pumping pulse and a direct detection of the refocused echo (b/ bottom spectrum). B0 is parallel to the c axis of 4H-SiC. In both PELDOR experiments, B0=3297 G, T= 300K, and the probe microwave frequency is fs= 9.308 GHz, such that the low field EPR line of the V2 spins seen on fig.3 b/ is here resonantly excited at fs. The microwave pump frequency fp is varied in the range [9.15; 9.4] GHz, thus over 250 MHz, in 250 steps of 1 MHz. Inset: pulses sequences applied: pink: optical pumping (first arrow); red: microwave pulse at fs (second arrow) and fp (third arrow); dark: the spin echo used for direct detection, ie the one integrated.} 
\end{figure}

The two obtained PELDOR spectrum correspond to the one expected. The preparatory experiment was the field sweep spectrum of fig.3 b/, thus the PELDOR spectrum (versus fp) should reflect somehow this field sweep spectrum (versus B0), but with of course an inverted order of the resonance lines. Of course, when fp=fs=9.308 GHz, the V2 spins feel an accelerated driven decoherence due to themselves, such that the probing EPR line chosen is always seen in such a PEDLOR spectrum versus fp. However, one can also distinguished on both PELDOR spectrum another dip occuring at fp= 9.243 GHz. This resonance line is 65 MHz below fs, corresponding to a resonnace line 23.2 G above the low field EPR line of the V2 shown on the field sweep spectrum of fig.3 b/. This line was thus also present on fig.3 b/ and correspond to other kinds of intrinsic defects in 4H-SiC having a g factor also very close to the one of the V2, that is close to g=2.0028. This line is often attributed to carbon related defects in bulk 4H-SiC. That means that the dip seen at fp= 9.243 GHz corresponds to the additionnal decoherence effect felt by the V2 spins probes due to the microwave driven manipulation of the carbon related defects located nearby them, producing through dipolar couplings, a fluctuating local magnetic field on the sites of the V2 spins probes. This is exactly the principle at the heart of quantum sensing as explained in my previous theoretical work on the SiC YIG quantum sensor~\cite{Tribollet2019dec}. The main difference is that here the target spin bath is 3D, whereas in ODPELDOR spectroscopy applied to structural biology, the target spin bath is 2D. Here also, direct detection is used instead of the highly sensitive photoluminescence detection.\\As a last remark, one notes that the third EPR line seen on the field sweep spectrum of fig.3 b/ is not seen here in the PELDOR spectrum. The reason is assumed to be the limited bandwidth available in the presented PELDOR experiments performed with a standard MS3 flexline resonator from Bruker. The MS3 cavity is known to have a bandwidth at half maximum of its microwave reflexion curve nearly equal to 100 MHz. As the central frequency of the cavity is here set equal to the resonant frequency of the V2 spins probes at fs=9.308 GHz under B0=3297 G, then any other resonant EPR line located much beyond +/- 50 MHz from fs=9.308 GHz can not be observed in the PELDOR spectrum by lack of microwave power at the associated pump frequency entering the cavity at those frequencies (the microwave power is reflected). The high field EPR line of the V2 spins occurs at B0= 3345 G on fig.3 b/ and should thus appears at 9243-65= 9178 MHz, which is thus not possible here. Those two room temperature optical pumping assisted PELDOR experiments thus: i/ clearly show that V2 spins probes, which are photoluminescent and which can thus be optically detected in an ultra sensitive manner by ODMR, can sense by microwave driven decoherence effects some paramagnetic centers located nearby them which are themselves not photoluminescent, thus providing a way to considerably increase the sensitivity of standard pulsed EPR spectrometers if the proposed SiC-YIG quantum sensor can be fabricated; ii/ they also demonstrate that using one single V2 EPR line and targeting a spin label EPR line located nearby the V2 EPR line, it should be clearly possible to perform ODPELDOR spectroscopy applied to structural biology with the SiC-YIG quantum sensor I previously proposed~\cite{Tribollet2019dec}.

\section{Spin wave resonance experiments on model ferromagnetic nanostripes of Permalloy}

Finally, I present in this section some test  spin wave resonance experiments, performed not on YIG nanostripes at X band, but on the more easily accessible Permalloy nanostripes at Q band (34 GHz). Those experiments were numerically simulated following the theoretical approach I previously presented in the context of quantum computing with an array of spin qubits in SiC located nearby a permalloy ferromagnetic nanostripe~\cite{Tribollet2014}.              

\begin{figure}[ht]             
\centering \includegraphics[width=1.1 \textwidth]{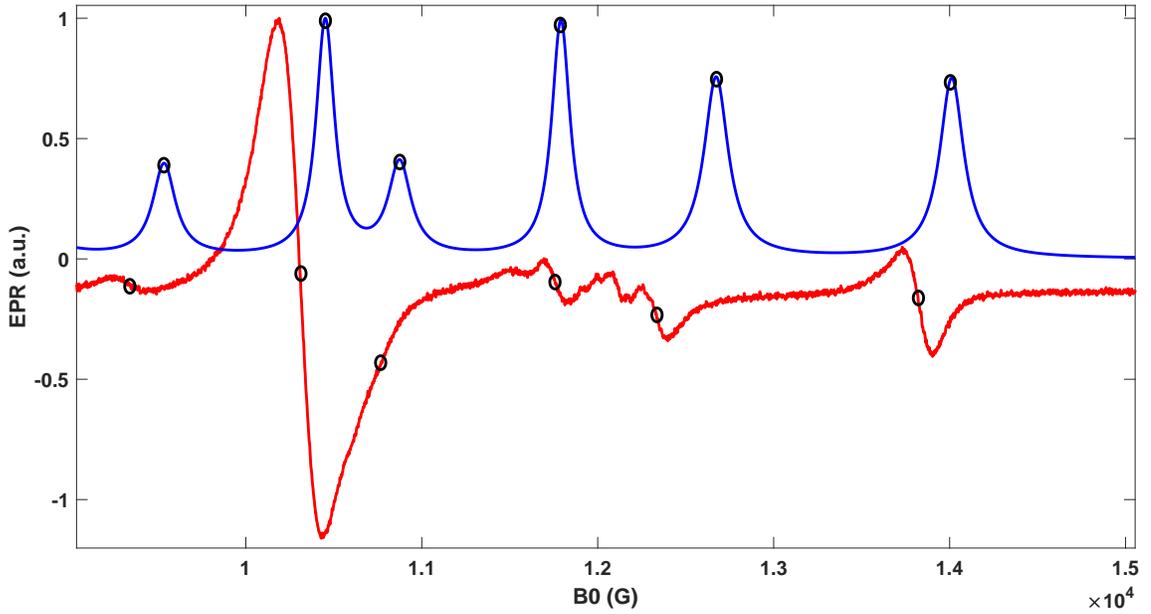}              
\caption{\label{fig_10} Spin Wave Resonance (SWR) spectrum of an ensemble of Permalloy ferromagnetic nanostripes (thickness T=100 nm, width w= 300 nm, and length $L\:=\:100\:\mu\!m$): in red: derivative spectrum, as measured at Q band (f= 34 GHz) with a magnetic field applied in the plane of the nanostripes, along the width w; in blue: absorption spectrum, as numerically simulated without any free parameter (and without considering the different oscillator strengths of the various SWR).}                                     
\end{figure}                        

For the numerical simulation, I used the saturation magnetization (11700 G) and g factor (2.00) known for Permalloy and the dimension of the Py nanostripes (thickness T=100 nm, width w= 300 nm, and length $L\:=\:100\:\mu\!m$). As it can be seen on fig.7, the experimentally observed spin wave resonance spectrum and the theoretical one match quite well, the six main spin wave resonance being obtained and having resonant magnetic fields close to the experimental ones, with an error on the order of one or few spin wave resonance linewidth. This is quite satisfactory considering the fact that there are no free parameter in the theoretical simulation presented here.

\section{Conclusion}
In this article, I have presented my first experiments towards the development of a new SiC-YiG hybrid quantum sensor compatible with a standard X band pulsed EPR spectrometer widely used worldwide. The measured T2 spin coherence time of $48\:\mu\!s$ at room temperature found here for bulk V2 spins in 4H-SiC confirm the high potential of those solid state spin qubits for quantum sensing application. My successful optically detected magnetic resonance experiments, as well as my optical pumping assisted PELDOR experiments confirm the relevance of this new experimental approach for state of art quantum sensing at the single spin sensitivity and with sub nanoscale resolution, interfacing a standard photoluminescence setup with a standrad pulsed ELDOR spectrometer by means of an optical fiber. This experimental approach and the related SiC-YIG quantum sensor should thus be of great interest for all the biophysicists, chemists and physicists which are already worldwide pulsed EPR user. The next challenges towards the practical demonstration of such a new experimental approach for state of art quantum sensing are i/ the fabrication by ion implantation of sub-surface quantum coherent isolated V2 spin probes, ii/ the fabrication of 4H-SiC nanophotonic structures for the efficient excitation and collection of V2 spin probes photoluminescence, and iii/ the fabrication of appropriate YIG nanomagnonic structures for the investigation of the depth profile of sub surface V2 color center spins created by ion implantation.\\ 
 
\section{Acknowledgments}       

The author thanks the University of Strasbourg and CNRS for the reccurent research fundings. The author also thanks the STNano central of technology in Strasbourg for fabricating and providing the model permalloy nanostripes studied here. 


\section{Competing financial interests}
The author declare that he has no competing financial interests.

\end{document}